\documentclass[preprint,pra,showpacs]{revtex4} 
 
\begin{document} 
\input{epsf}

\title{Effects of Vacuum Fluctuation Suppression \\
on Atomic Decay Rates}
\author{ L.H. Ford}
 \email[Email: ]{ford@cosmos.phy.tufts.edu} 
 \affiliation{Institute of Cosmology, 
Department of Physics and Astronomy\\ 
         Tufts University, Medford, MA 02155}
\author{Thomas A. Roman}
  \email[Email: ]{roman@ccsu.edu}
  \affiliation{Department of Mathematical Sciences \\
 Central Connecticut State University \\  
New Britain, CT 06050}

\begin{abstract}
The use of atomic decay rates as a probe of sub-vacuum phenomena
will be studied. Because electromagnetic vacuum fluctuations are
essential for radiative decay of excited atomic states, 
decay rates can serve as a measure of the suppression of
vacuum fluctuation in non-classical states, such as squeezed 
vacuum states. In such states the renormalized expectation value of the 
square of the electric field or the energy density
can be periodically negative, representing suppression of 
vacuum fluctuations. We explore the extent to which atomic
decays can be used to measure the mean squared electric field 
or energy density. We consider a scheme in which atoms in an
excited state transit a closed cavity whose lowest mode contains
photons in a non-classical state. The change in the decay probability
of the atom in the cavity due to the non-classical state can,
under certain circumstances, serve as a measure of the mean 
squared electric field or energy density in the cavity. We derive
a quantum inequality bound on the decrease in this probability.
We also show that the decrease in decay rate can sometimes be
a measure of negative energy density or negative squared
electric field. We make some estimates of the magnitude of this
effect, which indicate that an experimental test might be possible.
\end{abstract}

\maketitle

\baselineskip=14pt

\section{Introduction}
\label{sec:intro}
It has been known for some time that negative energy densities and
fluxes are a generic  prediction of quantum field theory~\cite{EGJ}. 
States which involve negative energy,  such as the Casimir vacuum
state and squeezed states of light,  have even been produced in the 
laboratory~\cite{C,SY}. Negative energy density may be viewed as an
example of a sub-vacuum phenomenon, whereby the vacuum fluctuations
are suppressed below their level in the Minkowski vacuum state. The
experiments which have been done measure an indirect effect, such
as a Casimir force or a change in photon counting statistics, not the
energy density itself. This raises the question of whether more direct
detection of negative energy density or related effects is possible.

The gravitational effects are far too small to be feasible in a
laboratory experiment. Furthermore, the magnitude and duration of
negative energy densities and fluxes are constrained by quantum
inequalities~\cite{F78,F91,LF100,TRMGM,CJF}. These  are 
constraints derivable directly from quantum field theory, which
yield an inverse relation between the magnitude of the negative
energy density and its duration. Marecki~\cite{M} proved quantum 
inequality-type bounds on the magnitude and duration 
of squeezing in quantum optics experiments involving squeezed states of light.
These constraints make the search for local measurements of sub-vacuum
effects more challenging.

Nonetheless, there seems to be no barrier in principle to constructing
a negative energy detector which relies solely upon electromagnetic
interactions. One model for such a detector was proposed by Ford,
Grove, and Ottewill~\cite{FGO}, who  analyzed a system of atomic spins, placed in 
an external magnetic field and coupled to a quantized electromagnetic
field. When the system is coupled to a non-classical state of photons,
such as a squeezed vacuum state, there can be a transient increase in
the average magnetic moment, as compared to when the quantized field
is in the vacuum state. This can be viewed as a suppression of a
depolarizing effect of vacuum fluctuations on the spins, 
resulting in momentary
``re-polarization''.  In general, this change is not directly 
correlated with energy density. However, for a certain ranges of parameters of the 
system, the change in magnetic moment is in phase with the periods of negative energy density.  
Under these circumstances, this spin system represents a non-gravitational negative energy detector. 
Unfortunately, the fractional change in the mean magnetic moment is
quite small unless the photon energies approach the $\gamma$-ray
range. Consequently, it is questionable that one could find a way
to detect the extremely rapid changes in the mean magnetic moment.

Other models have been treated by  Davies and Ottewill~\cite{DO}, who 
studied the detection of negative energy fluxes using various types of switched monopole 
particle detectors, building on earlier work by Grove~\cite{Grove}.
Marecki and Szpak \cite{MS} modeled spontaneous light emission 
from two-level atoms coupled to a quantized electromagnetic field. They derived a Volterra-type equation 
which controls the time evolution of the amplitude of the excited state.

In this paper, we follow the spirit of these earlier works and look for an indirect ``tracker''  
which might lead to the detection of negative energy or vacuum
fluctuation suppression-type effects. It is well-known that
electromagnetic vacuum fluctuations are essential for the spontaneous
decay of excited states of atoms. Without the coupling to the
quantized radiation field, all energy levels of an atom would be
eigenstates of the Hamiltonian, and hence stable. This suggests
that a suppression of the usual vacuum fluctuations could increase
the lifetime of an excited state in a way which might be observable.
 
We consider a model of a two-level excited atom interacting with a quantized electromagnetic field, using 
first-order perturbation theory. The field is confined to a closed cavity with two of its dimensions 
much larger than the other one. This is so that we can arrange the transit time of the atom through the cavity 
to be small compared to the light travel time across the larger two dimensions of the cavity. 
Our idea is to prepare the cavity field in a non-classical state, such as a squeezed 
vacuum, and fire the excited atom through the cavity along its shortest dimension, before the state 
of the field has time to change very much. We want the atom to interact with the field 
during the period when the renormalized expectation value of the square of the electric field, 
$\langle E^2 \rangle$, is negative. The purpose is to see whether this will suppress the de-excitation 
probability of the atom, compared to its vacuum value, in a way that will be correlated with the periods of 
$\langle E^2 \rangle<0$. If so, our system would function as a negative $E^2$ detector.  
We work primarily with cavity modes, for which $\langle E^2 \rangle \neq \langle B^2 \rangle$,  
so periods of $\langle E^2 \rangle<0$ do not generally
correspond to periods of negative energy density. However, under
certain conditions, our model will also serve to measure the energy density.
 
For most treatments of this type in quantum optics, the Jaynes-Cummings model for the 
interaction between the atom and the field is used. This model makes use of
the rotating wave approximation,
 which effectively ignores terms with rapidly oscillating exponentials~\cite{GK}. We explicitly 
do not  make this approximation, because the effects we are interested in are highly transient ones, 
and depend on these terms.

It also should be mentioned that any scheme which is designed so that a particle or observer 
interacts only with negative energy in flat spacetime can be ruled out by the 
quantum inequalities and the averaged weak energy condition, which is known to hold in 
Minkowski spacetime. In a realistic version of our atom-cavity system, the atom would have to pass through 
holes cut in the cavity walls. Even in the event that the atom passed through net 
negative energy while passing through the cavity,  
edge effects from the holes will contribute enough positive energy to satisfy 
the averaged weak energy condition. This effect has been discussed
recently by Graham and Olum~\cite{GO}, and  by Fewster, Olum, and
Pfenning~\cite{FOP}.

In our case,  we are comparing the de-excitation probability of the atom in an 
excited state of the cavity to when there is only vacuum in the cavity.  In particular, we 
are interested in situations where this probability is suppressed relative to its vacuum value, and where 
the periods of suppression are in phase with periods of $\langle E^2
\rangle<0$. We will be concerned with changes in $\langle E^2 \rangle$
or the energy density
due to changes in the quantum state in the cavity. In this case, edge
effects due to the holes in the cavity will cancel out. It is possible
for the difference in the net energy seen by an observer to be
negative, as was shown by Borde, Ford, and Roman~\cite{BFR}.

Throughout this paper, we will regard $\langle E^2 \rangle$ and the
energy
density as being set to zero in the vacuum state of the cavity. Thus
we are concerned only with changes due to changing the quantum state
of the cavity, and not with Casimir-type effects due to the geometry
of the cavity. It is well known that the presence of boundaries can
make the energy density or $\langle E^2 \rangle$ smaller than in
empty space vacuum state. However, this effect on decay rates of atoms
is difficult to distinguish from the effects of the changes in mode
structure, such as the change from a continuous to a discrete
spectrum. The effects of cavity geometry on atomic decays has been 
extensively studied in recent decades, with an early treatment given by
Babiker and Barton~\cite{BB}. For a recent review, 
see for example Ref.~\cite{HR}.

The outline of this paper is as follows: In Sect.~\ref{sec:AT-INT}, we
develop some formalism for describing the interaction of cavity modes
with an atom. This is further developed in Sect.~\ref{sec:DE-EX}, where we
obtain
detailed expressions for the de-excitation probability for an atom
traversing a cavity with one mode excited. In Sect.~\ref{sec:NE}, 
we make some numerical estimates of the size of the effect and 
discuss the feasibility of observing it. Our results are
summarized in Sect.~\ref{summary}. 
In the Appendix, we derive a quantum inequality-type bound on the 
de-excitation probability. Unless stated otherwise, we work in Lorentz-Heaviside units 
where $\hbar = c =1$.

\section{Atom-Cavity Interaction}
\label{sec:AT-INT}

Consider a two-level atom with states $|\psi_1 \rangle$ and
$|\psi_2 \rangle$. If the atom passes through a cavity containing a quantized radiation field, 
then the final state of the system 
will in general be an entangled one. First, consider the case where
the initial photon state in the cavity 
is a single-mode number eigenstate, $|\gamma_i \rangle=| n \rangle$, and the atom is 
in the state $|\psi_1 \rangle$. Then the initial state of the system is 
\begin{equation}
|\Psi_i \rangle = | n \rangle \, | \psi_1 \rangle \,.
\end{equation}
If we consider the case where the interaction changes the photon number 
by at most one, then the most general final state is of the entangled form:
\begin{equation}
|\Psi_f \rangle = B_1 \, | n \rangle \, |\psi_1 \rangle   + B_2 \, | n+1 \rangle \, |\psi_2 \rangle  
+ B_3  \,  | n-1 \rangle \, |\psi_1 \rangle  + B_4 \, | n+1 \rangle \, |\psi_1 \rangle 
+B_5 \, | n \rangle \, |\psi_2 \rangle  + B_6 \, | n-1 \rangle \, |\psi_2 \rangle               \,. 
\end{equation}
Suppose we want the probability of finding the atom in state $|\psi_2 \rangle$, irrespective 
of the photon state. Then we should project $|\Psi_f \rangle $ onto the subspace where 
 $|\psi_{atom} \rangle= |\psi_2 \rangle $:
\begin{equation}
\langle \psi_2 |\Psi_f \rangle = B_2 \, | n+1 \rangle \, + B_5 \, | n \rangle  \, +B_6 \, | n-1 \rangle \,.
\end{equation} 
However, $B_5=0$, because the interaction term only connects photon 
number states which differ by $\pm 1$.

The interaction Hamiltonian for the atom and the field in the cavity, 
in the dipole approximation, is given by:
\begin{equation}
H' = -{\bf d} \cdot {\bf E}_S  \,
\end{equation}
where ${\bf d}$ is the dipole moment of the atom and ${\bf E}_S$ is the 
Schr{\"o}dinger picture electric field operator 
for the quantized cavity field, evaluated at the atom's position.  
The dipole approximation assumes that we can ignore the variation of the field across the size of the atom.
The Schr{\"o}dinger picture electric field is related to the
Heisenberg field $ {\bf E} ({\bf x}, t)$ by   ${\bf E}_S({\bf x}) =
{\bf E} ({\bf x}, 0)$. The Heisenberg field has the mode expansion 
\begin{equation}
 {\bf E} ({\bf x}, t) = \sum_{k \lambda} [a_{k \lambda} {\bf \hat{e}}_{k \lambda}  
\, f_{k \lambda} ({\bf x}) \,e^{-i \omega t} + 
{a^\dagger}_{k \lambda} {\bf \hat{e}}_{k \lambda}  \, {f}_{k \lambda} ({\bf x}) \,e^{i \omega t}] \,. \label{eq:Efield}
\end{equation}
Here $f_{k \lambda} ({\bf x})$ is the spatial part of the mode
function, which we take to be real, and ${\bf \hat{e}}_{k \lambda}$ is a linear
polarization vector.  

First order perturbation theory yields
\begin{eqnarray}
B_2 &=& - i  \int_{t_0}^{t_1} \, dt' \, \langle n+1,
\psi_2 | H' |  n, \psi_1 \rangle \,
 e^{i(\omega-\Delta \varepsilon) t'}  \nonumber \\
&=& i \sqrt{n+1}\, \int_{t_0}^{t_1} \, dt' \, {\hat e}_{k \lambda} \cdot \langle \psi_2 | {\bf d} | \psi_1 \rangle  \,
f_{k \lambda}({\bf x} (t')) \, \, e^{i(\omega-\Delta \varepsilon) t'}   \\
B_6 &=& - i \int_{t_0}^{t_1} \, dt' \, \langle n-1, \psi_2 | H' |  n,
\psi_1 \rangle \, 
e^{-i(\omega+\Delta \varepsilon) t'}  \nonumber \\
&=& i \sqrt{n}\, \int_{t_0}^{t_1} \, dt' \, {\hat e}_{k \lambda} \cdot \langle \psi_2 | {\bf d} | \psi_1 \rangle  \,
f_{k \lambda}({\bf x} (t')) \, \, e^{-i(\omega+\Delta \varepsilon) t'}   \,.
\end{eqnarray}
We assume that $| \psi_1 \rangle$ is the higher of the two energy
states, and let $\Delta \varepsilon > 0$ be the energy difference 
between the two atomic states.
The probability of finding the atom in $|\psi_2 \rangle$ is
\begin{equation}
P_2 = {|\langle \psi_2 |\Psi_f \rangle |}^2 = {|B_2|}^2+{|B_6|}^2 \,.
\end{equation}
Note that the contribution of $B_6$ would be ignored in the rotating
wave approximation.

Let us now consider the case of a general one-mode initial photon
state 
\begin{equation}
| \gamma_i \rangle = \sum_{n=0}^{\infty} \, c_n \, | n \rangle \,.
                \label{eq:init_state}
\end{equation}
 The most general final state of the atom-cavity system is 
\begin{equation}
 |\Psi_f \rangle = \sum_{n=0}^{\infty} \,\, (A_{1n} | n \rangle \, | \psi_1 \rangle + 
 A_{2n} | n \rangle \, | \psi_2 \rangle )\,.
\end{equation}
Now the projection of $| \Psi_f \rangle$ onto $| \psi_2 \rangle$ is the vector in the photon 
state space given by 
\begin{equation}
\langle \psi_2 | \Psi_f \rangle =  \sum_{n=0}^{\infty} \, A_{2n} \, | n \rangle \,,
\end{equation}
and the probability of finding $| \psi_2 \rangle$ is
\begin{equation}
P_2= {|\langle \psi_2 | \Psi_f \rangle|}^2 =  \sum_{n=0}^{\infty} \, {| A_{2n} |}^2 \, .
\end{equation}

Let $T$ be the transition matrix, so that $T_{fi} = \langle f | T | i \rangle$ is the amplitude to make 
a transition from state $| i \rangle$ to state $| f \rangle$. Then we have that
\begin{eqnarray}
A_{2m} &=& \langle m \, \psi_2 | T | \gamma_i \, \psi_1 \rangle  \nonumber \\
&=&  \sum_{n=0}^{\infty} \, c_n \, \langle m \, \psi_2 | T | n \, \psi_1 \rangle \,.
\label{eq:A2m_1}
\end{eqnarray} 
This latter matrix element between energy eigenstates is then given in first-order perturbation 
theory by an integral of a matrix element of $H'$.

\section{De-excitation Probability for an Atom in a Cavity}
\label{sec:DE-EX}

\begin{figure} 
\begin{center} 
\leavevmode\epsfysize=5cm\epsffile{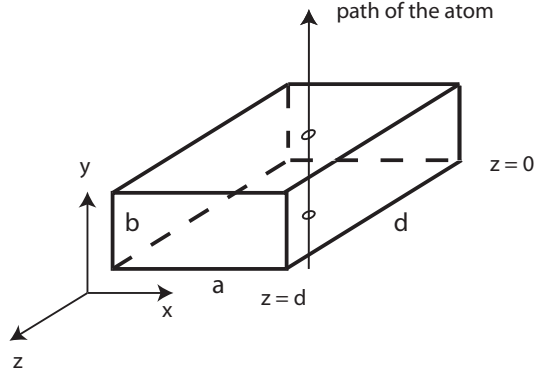}
\end{center} 
\caption{A cavity for standing wave modes. We require that $b<a<d$,
  and $b \ll d$. The atom is assumed to pass though the cavity in the
$y$-direction, as illustrated.}  
\label{fig:cavity} 
\end{figure}

Consider an atom passing through a rectangular cavity with dimensions
$a$, $b$, and $d$ aligned along the $x$, $y$, $z$ axes respectively, 
as shown in Fig.~\ref{fig:cavity}. 
We will assume that $b<a<d$, with $b \ll d$, and that the 
velocity of the atom is parallel to the $b$-dimension of the cavity,  
${\bf v} = v {\hat y}$. We will also
assume that $v \ll 1$, so that we may ignore relativistic effects.

\subsection{Perturbation Theory Results}

 Using first-order perturbation theory, we can write the transition 
matrix element between two states of definite photon number as
\begin{equation}
\langle m \psi_2 | T | n \psi_1 \rangle = 
- i \int_{t_0}^{t_1} \, dt' \, \langle m \psi_2 | H' | n \psi_1 \rangle
\,e^{i \Delta E_{sys} t'} \,,
\label{eq:T-melement1}
\end{equation}
where $ \Delta E_{sys}$ is the change in the energy of the atom-cavity system. The matrix element 
for the interaction Hamiltonian is
\begin{eqnarray}
 \langle m \psi_2 | H' | n \psi_1 \rangle & =& - \langle  \psi_2 | d_y |  \psi_1 \rangle \, 
  \langle m |{\hat E_y} | n  \rangle \nonumber \\
 &=&- \langle  \psi_2 | d_y |  \psi_1 \rangle \, \, (f \sqrt{n} \, \delta_{m,n-1} + 
 f \sqrt{n+1} \, \ \delta_{m,n+1} )\,.
\end{eqnarray}
Here we assume that the electric field in the cavity is polarized in
the $y$-direction.

The transition matrix element, Eq.~(\ref{eq:T-melement1}) can then be written as 
\begin{equation}
\langle m \psi_2 | T | n \psi_1 \rangle = i \,\langle  \psi_2 | d_y |  \psi_1 \rangle 
 (f \sqrt{n} \, \delta_{m,n-1} \, I_1+ 
 f \sqrt{n+1} \, \ \delta_{m,n+1}\, I_2 ) \,,
\end{equation}
where
\begin{equation}
I_1 =\int_{t_0}^{t_1} \, dt'  \,e^{-i (\omega+\Delta \varepsilon) t'} 
= \frac{e^{-i(\omega+\Delta \varepsilon) t_1} -e^{-i(\omega+\Delta \varepsilon) t_0} }
{-i (\omega+\Delta \varepsilon)} \,,
\label{eq:I_1}
\end{equation}
and
\begin{equation}
I_2 = \int_{t_0}^{t_1} \, dt'  \,e^{i (\omega-\Delta \varepsilon) t'}  
= \frac{e^{i(\omega-\Delta \varepsilon) t_1} -e^{i(\omega-\Delta \varepsilon) t_0} }
{i (\omega-\Delta \varepsilon)} \,.
\label{eq:I_2}
\end{equation}

Our Eq.~(\ref{eq:A2m_1}), assuming a general one-mode initial photon
state, Eq.~(\ref{eq:init_state}), becomes
\begin{equation}
A_{2m} = i \langle  \psi_2 | d_y |  \psi_1 \rangle \, [\sqrt{m+1} \, \,c_{m+1} \, f I_1
 + \sqrt{m}  \,\, c_{m-1} \, f I_2 ]\,.
\end{equation}  
The probability, $P_2$ of finding the atom in the lower energy state $|\psi_2 \rangle$ is 
\begin{eqnarray}
P_2 &=& \sum_{m} \, {|A_{2m}|}^2  \nonumber \\
& =& {|\langle  \psi_2 | d_y |  \psi_1 \rangle|}^2 \, 
\sum_{m=0}^{\infty} \Big[ (m+1) \, \, {|c_{m+1}|}^2 \, {|f|}^2 \, {| I_1 |}^2
 + m  \,\, {|c_{m-1}|}^2 \, {f}^2 {| I_2 |}^2 \nonumber \\
 &{ }&+
 2 Re \Big(\sqrt{m(m+1)} \, \,{f}^2  \,\,{I_1}^* I_2 \, \,{c^*}_{m+1} \,c_{m-1} \Big) \Big] \,.
\end{eqnarray}
We use the relations 
\begin{eqnarray}
\sum_{m} \, (m+1) \, {|c_{m+1}|}^2 &=& \langle n \rangle   \,, \nonumber \\
\sum_{m} \, m \, {|c_{m-1}|}^2 &=& \langle n \rangle +1  \,,  \nonumber \\
\sum_{m} \, \sqrt{m(m+1)} \, \, c_{m-1}\, {c^*}_{m+1} &=& \sum_{n} \, \sqrt{(n+1)(n+2)} \, c_n \, {c^*}_{n+2} \,.
\label{eq:sums}
\end{eqnarray}
Our expression for the probability is then
\begin{eqnarray}
P_2 & =&\, {|\langle  \psi_2 | d_y |  \psi_1 \rangle|}^2 \,
 \Big[ \langle n \rangle \,  {f}^2 \, {| I_1 |}^2
 + (\langle n \rangle+1)  \, \, {f}^2 {| I_2 |}^2 \nonumber \\
 &{ }&+
 2 Re \Big(\sum_{n=0}^{\infty} \, 
 \sqrt{(n+1)(n+2)} \, \, \, c_n \, {c^*}_{n+2} \,\,{f}^2  \,\,{I_1}^* I_2  \Big)\Big] \,.
\end{eqnarray}

In the case when the initial state of the field is the vacuum state, $|\gamma_i \rangle = |0 \rangle$, 
we have
\begin{equation}
P_2 = P_2(0) ={|\langle  \psi_2 | d_y |  \psi_1 \rangle|}^2 \, 
  \, {f}^2 {| I_2 |}^2 \,.
\end{equation}
It is important to note that this form of $P_2 (0)$ assumes that the atom decays into only 
one mode. This should be a good approximation near resonance, but not otherwise. Let us now consider the ratio
\begin{equation}
\frac{P_2}{P_2 (0)} = \langle n \rangle {\Bigg |\frac{I_1}{I_2}\Bigg |}^2 + \langle n \rangle +1 +
2 \sum_{n=0}^{\infty} \,  \sqrt{(n+1)(n+2)} \, \, \,Re \Bigg[ c_n \, {c^*}_{n+2} \,\,
 \frac{I_2 I^*_1}{{|I_2|}^2} \Bigg]\,.
\end{equation}
 {} From Eqs.~(\ref{eq:I_1}) and (\ref{eq:I_2}), we have that 
\begin{equation}
{\Bigg |\frac{I_1}{I_2}\Bigg |}^2 = \frac{1-{\rm cos}[(\omega+\Delta \varepsilon) (t_1-t_0)]}
{1-{\rm cos}[(\omega-\Delta \varepsilon) (t_1-t_0)]} \,\, \frac{{(\omega -\Delta \varepsilon)}^2}
{{{(\omega +\Delta \varepsilon)}}^2} \,.
\end{equation}
In the limit when the transit time of the atom through the cavity is very short, $t_1 \rightarrow t_0$, 
and we have
\begin{equation}
{\Bigg |\frac{I_1}{I_2}\Bigg |}^2 \approx 1 \,.
\end{equation}

In the limit when $\Delta \varepsilon \rightarrow \omega$, 
\begin{eqnarray}
{\Bigg |\frac{I_1}{I_2}\Bigg |}^2 &\approx& \frac{1-{\rm cos} \,2 \omega(t_1-t_0)}{2 \omega^2 {(t_1-t_0)}^2} 
\nonumber \\
I_2 I^*_1 &\approx& -\frac{i}{2} \,\frac{(t_1-t_0)}{\omega} \Big(e^{2 i \omega t_1}-e^{2 i \omega t_0} \Big) 
\nonumber \\
{|I_2|}^2  &\approx& {(t_1-t_0)}^2 \,,
\end{eqnarray}
and therefore we have that 
\begin{equation}
\frac{I_2 I^*_1}{{|I_2|}^2} \approx -\frac{i \, e^{2 i \omega t_0}}{2 \omega\,(t_1-t_0)} 
\Big[e^{2 i \omega (t_1-t_0)} - 1 \Big] \,.
\end{equation}
In addition to letting $\Delta \varepsilon \rightarrow \omega$, if we now also let $t_1 \rightarrow t_0$, 
the last equation above reduces to 
\begin{equation}
\frac{I_2 I^*_1}{{|I_2|}^2} \approx  -\frac{i \, e^{2 i \omega t_0}}{2 \omega\,(t_1-t_0)}  2 i \omega (t_1-t_0) 
= e^{2 i \omega t_0} \,.
\end{equation}
Thus if we assume $(t_1-t_0) \omega \ll 1$ and $(t_1-t_0) \Delta \varepsilon \ll 1$, then 
\begin{equation}
{\Bigg|\frac{I_1}{I_2} \Bigg|}^2 \approx 1 \,.
\end{equation}
If in addition, we take $\Delta \varepsilon \approx \omega$, we have
\begin{equation}
\frac{I_2 I^*_1}{{|I_2|}^2} \approx  e^{2 i \omega t_0} \,.
\end{equation}
Therefore, the ratio of probabilities becomes
\begin{equation}
\frac{P_2}{P_2 (0)} = 2 \langle n \rangle  +1 +2 \sum_{n=0}^{\infty} \,  \sqrt{(n+1)(n+2)} 
\, \, \,Re \Big( c_n \, {c^*}_{n+2} \,\,  e^{2 i \omega t_0} \Big) \,.
\label{eq:P-ratio-final}
\end{equation}

Let us compare this with the expression for the expectation value, $\langle {\hat E}^2 \rangle$, 
of the normal-ordered squared electric field operator in the cavity. We can 
calculate this using the expansion of the operator in
Eq.~(\ref{eq:Efield}) and the state given in Eq.~(\ref{eq:init_state}) for a single mode.
 We have
\begin{eqnarray}
   \langle  E^2 ({\bf x_0}, t) \rangle = 
  \langle \gamma_i|  { E}^2 ({\bf x_0}, t) | \gamma_i \rangle & =&
{f^2} ({\bf x_0}) \, \sum_{nl} \, {c^*}_n  \, c_l  \,\langle n  \, | 2 a^\dagger a + a^2 e^{-2i \omega t} 
+{(a^\dagger)}^2 e^{2 i \omega t} | \, l  \rangle \nonumber \\
&=& {f^2} ({\bf x_0}) \, \sum_{nl} \, {c^*}_n  \, c_l  \, \Big(2\, n \, \delta_{ln} + 
\sqrt{(n+2)(n+1)} \, \delta_{l,n+2} \,\, e^{-2 i \omega t}   \nonumber \\
&{}{}& + \sqrt{n (n-1)} \, \delta_{l,n-2} \,\, e^{2 i \omega t} \Big) \nonumber \\
&=& {f^2} ({\bf x_0}) \, \sum_{n} \, \Big[2 n {|c_n|}^2 + \sqrt{(n+2)(n+1)} \, \, {c^*}_n  \, c_{n+2} \,\, e^{-2i \omega t}  
\nonumber \\
&{}{}&+  \sqrt{n (n-1)} \,  \, {c^*}_n  \, c_{n-2} \,\, e^{2i \omega t} \Big] \,.
\label{eq:E^2}
\end{eqnarray}

If we relabel $n \rightarrow n+2$, we can rewrite the last sum in the following way:
\begin{eqnarray}
\sum_{n=2} \,\sqrt{n (n-1)} \,  \, {c^*}_n  \, c_{n-2} \,\, e^{2i \omega t}& =& 
\sum_{n=0} \,\sqrt{(n+2) (n+1)} \,  \, {c^*}_{n+2}  \, c_{n} \,\, e^{2i \omega t} \nonumber \\
&=&\sum_{n=0} \,\sqrt{(n+2) (n+1)} \,  \, {( {c}_{n+2}  \, {c^*}_{n} \,\, e^{-2i \omega t})}^* \,.
\end{eqnarray}
Using this and the fact that 
\begin{equation}
\sum_{n=0}  \, n\, {|c_n|}^2  = \langle n \rangle \,,
\end{equation}
we may write Eq.~(\ref{eq:E^2}) as 
\begin{equation}
\langle  E^2 ({\bf x_0}, t) \rangle  =
 {f^2} ({\bf x_0})\,\Big[ \, 2 \,  \langle n \rangle \,
+ 2 \sum_{n=0}  \sqrt{(n+2) (n+1)} \,  \,  Re(c_n \, {c^*}_{n+2} \, e^{2i \omega t}) \Big] \,.
\label{eq:expE2}
\end{equation}
Near resonance, i.e., in the limit  $\Delta \varepsilon \rightarrow \omega$, 
and in the limit of short transit times for the atom, i.e.,  $t_1 \rightarrow t_0$, 
the ratio of the de-excitation probabilities, ${P_2}/{P_2 (0)} $, can therefore be written in terms of 
$\langle  E^2 ({\bf x_0}, t) \rangle$ as 
\begin{equation}
\frac{P_2}{P_2 (0)} =1+ \frac{1}{ {f^2} ({\bf x_0})} \, 
 \langle  E^2 ({\bf x_0}, t) \rangle  \,.
\label{eq:P2/P20_1}
\end{equation}
Therefore we see that $P_2/P_2 (0)<1$ when $\langle E^2 \rangle < 0$. In the Appendix, we show 
that $\langle E^2 \rangle $ is bounded from below by $-f^2({\bf
  x_0})$. This guarantees that  $P_2/P_2 (0)$ is non-negative, as required.

Next we consider the case where we are near resonance, $\Delta \varepsilon \rightarrow \omega$, 
but with no restriction on $t_1-t_0$. The ratio of probabilities is
\begin{eqnarray}
\frac{P_2}{P_2 (0)} = \langle n \rangle \, {\Bigg |\frac{I_1}{I_2} \Bigg |}^2 + \langle n \rangle +1 
+2 \sum_{n}  \sqrt{(n+1) (n+2)} \,  \,  Re \Bigg(c_n \, {c^*}_{n+2} \frac{{I^*}_1 I_2}{{|I_2|}^2}\Bigg)  \,,
\label{eq:P-ratio}
\end{eqnarray}
where 
\begin{eqnarray}
 {\Bigg |\frac{I_1}{I_2} \Bigg |}^2 &\approx&  \frac{1-{\rm cos} \,2 \omega(t_1-t_0)}{2 \omega^2 {(t_1-t_0)}^2} 
 \nonumber \\
 \frac{I_2 I^*_1}{{|I_2|}^2} &\approx& -\frac{i \, e^{2 i \omega t_0}}{2 \omega\,(t_1-t_0)} \,
\Big[e^{2 i \omega (t_1-t_0)} - 1 \Big] \,.
\end{eqnarray}
In this case, Eq.~(\ref{eq:P-ratio}) becomes
\begin{eqnarray}
\frac{P_2}{P_2 (0)} &=& \langle n \rangle +1 +  \langle n \rangle \, \,
\frac{1-{\rm cos} \,2 \omega(t_1-t_0)}{2 \omega^2 {(t_1-t_0)}^2}  \nonumber \\
&{}{}&- \sum_{n=0}^{\infty}  \frac{\sqrt{(n+1) (n+2)}}{ \omega \, (t_1-t_0)} \,  Re \Big[i \, c_n \, {c^*}_{n+2} 
\, \, e^{2i \omega t_0} \,\Big(e^{2 i \omega (t_1-t_0)} - 1 \Big) \Big]\,.
\end{eqnarray}
Note that $\langle E^2 \rangle$ contains the factor $e^{2 i \omega t}$ and the integral of this expression,
\begin{equation}
\int_{t_0}^{t_1} \, dt \, e^{2 i \omega t} = \frac{1}{2 i \omega} \,(e^{2 i \omega t_1} - e^{2 i \omega t_0})\,
\end{equation}
does not have a factor of $1/(t_1-t_0)$. Thus in general, ${P_2}/{P_2 (0)}$ does not seem to be 
proportional to either $\langle E^2 \rangle$ or its time integral. When $\omega(t_1-t_0) \gg 1$, we have 
\begin{equation}
\frac{P_2}{P_2 (0)} = \langle n \rangle +1 + O \Bigg(\frac{1}{\omega(t_1-t_0)} \Bigg ) 
+ \langle n \rangle \, O \Bigg(\frac{1}{\omega^2 {(t_1-t_0)}^2} \Bigg ) \,.
\end{equation}
Therefore, in this limit ${P_2}/{P_2 (0)} >1$.

The advantage of considering the ratio ${P_2}/{P_2 (0)} $ is that the
precise form of the mode functions does not matter, since they cancel
out. 
However, the drawback is that our expression, 
Eq.~(\ref{eq:P2/P20_1}),  is only valid near resonance 
since we assume decay into only one mode. To avoid this limitation, 
it is also useful to look at the difference, $\Delta P_2 = P_2 -
P_2(0)$, so that the contribution of 
the unexcited modes cancels, and hence we do not need to be near resonance. 
The difference, $\Delta P_2$, is given by
\begin{eqnarray}
\Delta P_2  & =& {|\langle  \psi_2 | d_y |  \psi_1 \rangle|}^2 \, {f}^2 \Big[\langle n \rangle 
\Big ({|I_1|}^2+{|I_2|}^2 \Big) \nonumber \\
&{}{}&+ \,2 \sum_{n}  \sqrt{(n+1) (n+2)} \,  \,  Re \Big(c_n \, {c^*}_{n+2} {I^*}_1 I_2 \Big)  \,,
\end{eqnarray}
where 
\begin{eqnarray}
{|I_1|}^2 & =&  2 \,\, \frac{1-{\rm cos} (\omega+\Delta \varepsilon) (t_1-t_0)}{{(\omega + \Delta \varepsilon)}^2} \,,
\nonumber \\
{|I_2|}^2 & =&  2 \,\,\frac{1-{\rm cos} (\omega-\Delta \varepsilon) (t_1-t_0)}{{(\omega - \Delta \varepsilon)}^2}\,,
\nonumber \\
{I^*}_1 I_2  &=& \frac{e^{2 i \omega t_1} +e^{2 i \omega t_0}
-e^{ i [\omega (t_1+t_0)+\Delta \varepsilon (t_1-t_0)]} - 
e^{ i [\omega (t_1+t_0)-\Delta \varepsilon (t_1-t_0)]}}{{\Delta \varepsilon}^2-\omega^2} \,.
\end{eqnarray} 
We have assumed that the mode functions, $f$, are real. However this assumption 
is not really necessary, since one can always absorb the phase of the $f$'s into the phases of the 
complex coefficients, i.e., the $c$'s. 

In the case where we are far below resonance, $\omega \ll \Delta \varepsilon$,
\begin{eqnarray}
{|I_1|}^2  \approx {|I_2|}^2  &\approx& \,\,  \frac{2}{{ \Delta \varepsilon}^2} \,
[1-{\rm cos} \,\Delta \varepsilon (t_1-t_0)]\,,
\nonumber \\
{I^*}_1 I_2 & \approx& \frac{[2-e^{ i \Delta \varepsilon (t_1-t_0)} - 
e^{- i \Delta \varepsilon (t_1-t_0)}]}{{\Delta \varepsilon}^2} 
\nonumber \\
& \approx& \frac{2}{{ \Delta \varepsilon}^2} [1-{\rm cos} \,\Delta \varepsilon (t_1-t_0)] \,.
\label{eq:Is}
\end{eqnarray}
This assumes that $\omega t_0$ and $\omega t_1$ are both much less
than 1. Physically, we  require that $\omega (t_1-t_0) \ll 1$, 
and then for convenience set $t_0=0$. Using the above expressions,  
$\Delta P_2 $ becomes 
\begin{eqnarray}
\Delta P_2 &\approx &4 \, {|\langle  \psi_2 | d_y |  \psi_1 \rangle|}^2 \, {f}^2
\frac{ [1-{\rm cos} \,\Delta \varepsilon (t_1-t_0)]}{{ \Delta \varepsilon}^2} \, 
\times \Big[\langle n \rangle 
+\sum_{n}  \sqrt{(n+1) (n+2)} \,  \,  Re \Big(c_n \, {c^*}_{n+2} \Big) \Big]   \nonumber \\
&\approx & 2 \, {|\langle  \psi_2 | d_y |  \psi_1 \rangle|}^2 \, 
\frac{ [1-{\rm cos} \,\Delta \varepsilon (t_1-t_0)]}{{ \Delta \varepsilon}^2} \,  
\langle  {\hat E}^2 ({\bf x_0}, t_0)\rangle  \,.
\label{eq:DeltaP2}
\end{eqnarray}
Thus we can have $\Delta P_2  < 0$, if $\langle E^2 \rangle < 0$ at the time that the atom transits the cavity. 
In the Appendix we show that $\langle E^2 \rangle $ is bounded from below and that, as a result, we 
have Eq.~(\ref{eq:DeltaP2_QI}) 
\begin{equation}
\Delta P_2 
\geq - \frac{2 \, {|\langle  \psi_2 | d_y |  \psi_1 \rangle|}^2 \, {f^2} ({\bf x_0}) }
{{ \Delta \varepsilon}^2} \,.
\label{eq:DeltaP2_QIbound}
\end{equation}
This result provides a limit on the degree to which sub-vacuum effects
may suppress the decay probability. This limit is analogous to the
quantum inequality bounds on negative energy densities and 
fluxes~\cite{F78,F91,LF100,TRMGM,CJF}.

In the case where we are far above the resonant frequency, 
$\omega \gg \Delta \varepsilon$ and when \hbox{$\Delta \varepsilon (t_1-t_0)
\ll 1$},
we have 
\begin{eqnarray}
{|I_1|}^2  \approx {|I_2|}^2  &\approx& \,\,  \frac{2}{ \omega^2} \,
[1-{\rm cos} \,\omega (t_1-t_0)]\,,
\nonumber \\
{I^*}_1 I_2 & \approx& -\frac{{(e^{ i \omega t_1} - 
e^{ i \omega t_0})}^2}{\omega^2}  \,,
\label{eq:Is_2}
\end{eqnarray}
and 
\begin{eqnarray}
\Delta P_2 & \approx& \frac{2}{\omega^2} \,{|\langle  \psi_2 | d_y |  \psi_1 \rangle|}^2 \, {|f|}^2 \,
\Big \{2 \langle n \rangle 
\Big( 1-{\rm cos} \,\omega (t_1-t_0) \Big) \nonumber \\
&{}{}&-  \sum_{n}  \sqrt{(n+1) (n+2)} \,  \,  Re \Big[c_n \, {c^*}_{n+2} {(e^{ i \omega t_1} - 
e^{ i \omega t_0})}^2 \Big)  \Big] \Big\}\,.
\end{eqnarray}
Although it is possible to have $\Delta P_2 <0$, periods of $\Delta P_2 <0$ do not seem to be correlated 
with either $\langle E^2 \rangle$ or its time integral in this case.

\subsection{Specific quantum states}
\label{sec:states}

In this subsection, we will discuss two specific non-classical  states
of the photon field.

\subsubsection{Vacuum plus two photon state}
\label{sec:0p2}

\begin{figure} 
\begin{center} 
\leavevmode\epsfysize=6cm\epsffile{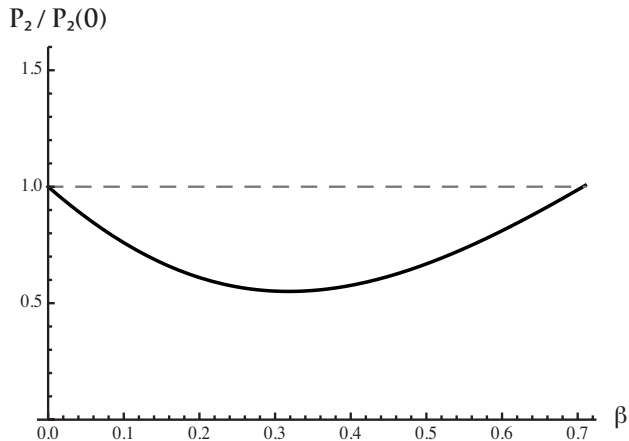} 
\end{center}
\caption{The ratio of the decay probability in the vacuum plus two
  photon state to that in the vacuum state is plotted as a function
of the parameter $\beta$.}  
\label{fig:ratio} 
\end{figure}

One of the simplest examples of a quantum state which exhibits negative
energy density and $\langle E^2 \rangle < 0$ is a coherent
superposition of the vacuum and a state containing two photons in the
same mode. Such a state can be expressed as
\begin{equation}
|\gamma \rangle = \frac{1}{\sqrt{1+\beta^2}} \;(|0\rangle + \beta
  |2\rangle ) \,,
\label{eq:02}
\end{equation}
where we take $\beta$ to be a real, non-negative parameter. In this
state, the mean squared electric field may be written as
\begin{equation}
 \langle  E^2 ({\bf x_0}, t) \rangle = \frac{2\beta}{1+\beta^2}\,
 {f^2} ({\bf x_0})\, [2 \beta +\sqrt{2}\, \cos(2\omega t)] \,,
\label{eq:02Esq}
\end{equation}
which reaches its minimum value when $ \cos(2\omega t) =-1$. If
$\beta < \sqrt{2}/2$, then this value is negative. At this point, the
ratio of probabilities may be expressed as
\begin{equation}
\frac{P_2}{P_2 (0)} = 1 +  \frac{2\beta}{1+\beta^2}\;
(2 \beta +\sqrt{2}) \,.
\label{eq:02ratio}
\end{equation}
This ratio is plotted in Fig.~\ref{fig:ratio}, from which we can see
that ${P_2}/{P_2 (0)}$ reaches its minimum value of about $0.55$ when
$\beta \approx 0.32$. Thus it is possible to achieve a $45\%$ reduction
in the decay probability in this state.

\begin{figure} 
\begin{center} 
\leavevmode\epsfysize=6cm\epsffile{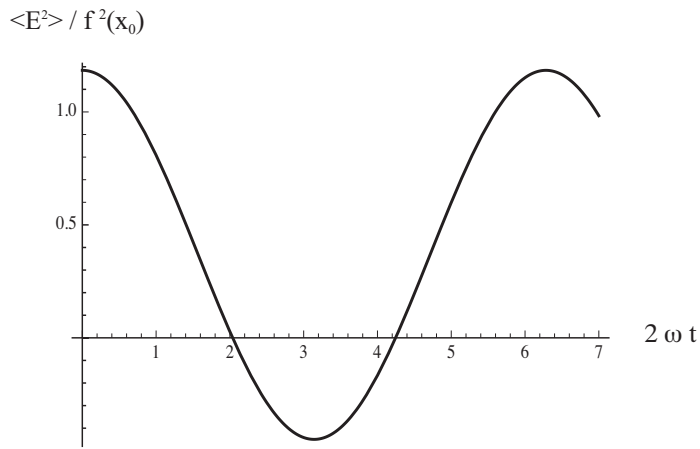} 
\end{center}
\caption{The mean squared electric field in the vacuum plus two photon
state is plotted as a function of time for the case $\beta = 0.32$.}  
\label{fig:E2} 
\end{figure}

The mean squared electric field is plotted as a function of time 
in Fig.~\ref{fig:E2} for the case $\beta = 0.32$. Note that
 $\langle E^2 \rangle < 0$ about $1/3$ of the time. However, because
 $\langle E^2 \rangle$ oscillates at an angular frequency of $2\omega$,
the duration of the interval when  $\langle E^2 \rangle < 0$ is
approximately $\Delta t \approx \pi/(3 \omega) \approx 1/\omega$.

\subsubsection{Squeezed vacuum state}

Another example of a quantum state which exhibits sub-vacuum effects is
the squeezed vacuum state, described by the complex parameter $\zeta =
r e^{i\phi}$. This state has been studied extensively, beginning with
the work of Caves~\cite{Caves}. Using the results in
Ref.~\cite{Caves}, one may show that the mean squared electric field
in a squeezed vacuum state becomes
\begin{equation}
\langle  E^2 ({\bf x_0}, t) \rangle = 2 {f^2} ({\bf x_0})\,
[\sinh^2 r + \cosh r \, \sinh r \cos (\phi + 2\omega t)]\,.
\label{eq:sqE2}
\end{equation}
This quantity is most negative when $ \cos (\phi + 2\omega t) = -1$,
at which point we have
\begin{equation}
\langle  E^2 ({\bf x_0}, t) \rangle = -{f^2} ({\bf x_0})\,(1 - e^{-2r})\,.
\label{eq:sqE2_2}
\end{equation}
If we compare this relation with Eq.~(\ref{eq:P2/P20_1}), we see that 
$P_2 \rightarrow 0$ for $r \gg 1$. Thus, in the limit of large squeeze
parameter, the decay rate can momentarily go close to zero. It is also of
interest to note that in this limit, the inequality
Eq.~(\ref{eq:DeltaP2_QIbound}) becomes an equality, as may
be seen from Eqs.~(\ref{eq:DeltaP2}) and (\ref{eq:E2bound}).

\subsection{Standing wave modes}
\label{sec:cavity modes}
To make the time interval when $\langle E^2 \rangle < 0$ as long as possible, we are interested 
in the lowest frequency mode of the cavity depicted in Fig. 1. A straightforward calculation 
using the formalism of Chapter 8 of Jackson \cite{Jackson} shows that, with our condition that $b<a<d$, the 
lowest frequency mode is the TE mode with $p=l=1,m=0$, where
\begin{equation}
\omega = \pi \sqrt{\frac{1}{a^2}+\frac{1}{d^2}} \,,
\label{eq:lowestmode}
\end{equation}
and
\begin{eqnarray}
E_x &=&  E_z=0  \,,\nonumber\\
E_y&=& \frac{ \omega a}{\pi} \, A_{10} \, \sin \Big(\frac{\pi}{a} x \Big)\, 
\sin  \Big(\frac{\pi}{d} z \Big) \,, \nonumber \\
B_x &=& i\frac{a}{d} \, A_{10} \, \sin \Big(\frac{\pi}{a} x \Big)\, 
 \cos \Big(\frac{\pi}{d} z \Big) \,, \nonumber \\
B_y &=& 0 \,, \nonumber \\
B_z &=& -i A_{10} \,\cos \Big(\frac{\pi}{a} x \Big) \, \sin \Big(\frac{\pi}{d} z \Big) \,, 
\label{eq:EBs}
\end{eqnarray}
where the electric field is taken to be polarized in the $y$-direction.
Here $A_{10}$ is a real normalization constant, which we will now determine.

In Lorentz-Heaviside units, we can write the energy density for a classical electromagnetic  field as 
\begin{eqnarray}
\rho&=&\frac{1}{2} ( {\bf E}^2 + {\bf B}^2) = \frac{1}{2}( { E_y }^2 + {|B_x|}^2+ {|B_z|}^2) 
\nonumber \\
&=& \frac{1}{2} {A_{10}}^2 \, \Big[\frac{\omega^2 a^2}{\pi^2} \, {\rm sin}^2  \Big(\frac{\pi}{a}x \Big)\,
{\rm sin}^2 \Big(\frac{\pi}{d}z \Big) +\frac{a^2}{d^2} \, {\rm sin}^2 \Big(\frac{\pi}{a}x \Big)\,
{\rm cos}^2 \Big(\frac{\pi}{d}z \Big)  
\nonumber \\
&{}{}&+  {\rm cos}^2 \Big(\frac{\pi}{a}x \Big)\,
{\rm sin}^2 \Big(\frac{\pi}{d}z \Big) \Big] \,, 
\label{eq:rho-class}
\end{eqnarray}
where we have used the expressions given in Eq.~(\ref{eq:EBs}) above. 
Let us normalize the vacuum mode functions by setting 
\begin{equation}
\int d^3x \, \rho = \int_{0}^a dx \,\int_{0}^b dy  \, \int_{0}^d dz \, \rho = \frac{1}{2} \omega \,.
\end{equation}
This leads to the result
\begin{equation}
{A_{10}}^2 = \frac{2 \omega}{a\,b\,d\, \Big(1+\frac{a^2}{d^2} \Big)} \,,
\end{equation}
and
\begin{equation}
{E_y}^2 = \frac{2 \omega^3 \, a}{\pi^2 \,b\,d\, \Big(1+\frac{a^2}{d^2} \Big)} \, 
{\rm sin}^2  \Big(\frac{\pi}{a}x \Big)\,{\rm sin}^2 \Big(\frac{\pi}{d}z \Big) \,.
\label{eq:Ey2}
\end{equation}

\subsection{Difference in probabilities for low $\omega$}
\label{sec:ProbDiff}
We may use the above results to give expressions for the quantity
$\Delta P_2$ in the case $\omega \ll \Delta \varepsilon$. Combining 
Eq.~(\ref{eq:DeltaP2}), and Eqs.~(\ref{eq:Ey2}) and
(\ref{eq:lowestmode}) 
for the $p=l=1,m=0$, TE mode, we have
\begin{eqnarray}
\Delta P_2 
&\approx&   8 \pi \, {\Bigg(1+\frac{a^2}{d^2} \Bigg)}^{1/2}
\frac{  {|\langle  \psi_2 | d_y |  \psi_1 \rangle|}^2 \, }
{ a^2 \,b\,d  \,{ ( \Delta \varepsilon)}^2} \, 
\, {\rm sin}^2  \Big(\frac{\pi}{a}x \Big)\,{\rm sin}^2 \Big(\frac{\pi}{d}z \Big) 
\nonumber \\
&{}{}{}& \times [1-{\rm cos} \,\Delta \varepsilon (t_1-t_0)] \, \times \Big[\langle n \rangle 
+\sum_{n}  \sqrt{(n+1) (n+2)} \,  
 \,  Re \Big(c_n \, {c^*}_{n+2} \Big) \Big] \, .
\label{eq:DeltaP2_2}
\end{eqnarray}

\subsection{Does $\langle E^2 \rangle < 0$ imply $\langle \rho \rangle < 0$?}
\label{sec:E^2vsrho}
Ideally one would like our system to be such that $\langle E^2 \rangle < 0$ 
when $\langle \rho \rangle < 0$, so that $\langle E^2 \rangle $  tracks negative energy 
density. For a classical single plane wave mode, $E^2 =B^2$, in Lorentz-Heaviside units. 
Therefore, for the quantized field, $\langle E^2 \rangle=\langle B^2 \rangle$, and 
$\rho = (\langle E^2 \rangle+\langle B^2 \rangle)/2$ imply that $\langle E^2 \rangle < 0$ 
when $\langle \rho \rangle < 0$. Unfortunately, for modes in a cavity the situation is more 
complicated, and in general $\langle E^2 \rangle \neq \langle B^2 \rangle$.
We may rewrite Eq.~(\ref{eq:Efield}) as
\begin{equation}
{\bf E} = \sum_{j}  \, (a_j  \,  {\bf E_j} + {a_j}^{\dagger} \,  {\bf
  E_j}^*) \, ,
\end{equation}
and the corresponding expression for the magnetic field operator as
\begin{equation}
{\bf B} = \sum_{j}  \, (a_j  \,  {\bf B_j} + {a_j}^{\dagger} \,  {\bf
  B_j}^*) \, .
\end{equation}
Here
\begin{equation}
 {\bf E_j}= {\hat {\bf e}}_j \, f_j ({\bf x}) \,e^{-i \omega t}
\end{equation}
and 
\begin{equation}
 {\bf B_j}= \frac{i}{\omega}\, 
{\hat {\bf e}}_j {\bf \times \nabla} f_j ({\bf x}) \,e^{-i \omega t}\,.
\end{equation}
The latter expression follows from the Maxwell equation ${\bf \nabla
  \times E} = - {\bf \dot{B}}$, and the vector identity
 ${\bf \nabla  \times}({\hat {\bf e}} f)= 
{\hat {\bf e}}{\bf \times \nabla} f$ for a constant vector
${\hat {\bf e}}$.

For the case where only a single mode $j$ is excited,  the normal-ordered 
expectation values of the squared fields are
\begin{equation}
\langle E^2 \rangle = 2 \langle a^\dagger \, a\rangle  \, |{\bf E}_j|^2+ 
2Re \Big( \langle  a^2 \rangle \, {\bf E}_j^2 \Big) 
\end{equation}
and 
\begin{equation}
\langle B^2 \rangle = 2 \langle a^\dagger \, a\rangle  \, |{\bf
  B}_j|^2 + 
2Re \Big( \langle  a^2 \rangle \, {\bf B}_j^2 \Big)\,.
\end{equation}
For a general mode function, the second terms in the expressions for $\langle E^2 \rangle$ 
and $\langle B^2 \rangle$ are not in phase.  

For example, consider our situation of interest, a single TE mode in a rectangular cavity, with 
$p=l = 1, m=0$. For this mode we have, from Eq.~(\ref{eq:EBs}), that 
\begin{eqnarray}
 {\bf E}_j^2& =& {E_y}^2 =  \frac{\omega^2 a^2}{\pi^2} \,{A_{10}}^2 \, 
 {\rm sin}^2  \Big(\frac{\pi}{a}x \Big)\, {\rm sin}^2
 \Big(\frac{\pi}{d}z \Big)\, e^{-2i\omega t}
\nonumber \\
 {\bf B}_j^2 &=& B_x^2 + B_z^2 
= - {A_{10}}^2 \Bigg[{\rm cos}^2 \Big(\frac{\pi}{a}x \Big) \, {\rm sin}^2 \Big(\frac{\pi}{d}z \Big)
+ \frac{a^2}{d^2} \, {\rm sin}^2 \Big(\frac{\pi}{a}x \Big)\,
{\rm cos}^2 \Big(\frac{\pi}{d}z \Big)   \Bigg]\, e^{-2i\omega t} \,.
\end{eqnarray}
These two expressions have opposite signs, so when $\langle E^2 \rangle<0$, $\langle B^2 \rangle>0$, 
and vice versa. Therefore,  periods when $\langle E^2 \rangle<0$ do 
not, in general, necessarily correspond to periods of negative energy
density. 

However, there are special cases when $|\langle E^2 \rangle|
\gg |\langle B^2 \rangle|$ and $\rho \approx \langle E^2 \rangle$. For
the mode discussed above, this occurs when $x \approx a/2$ and either
$z \approx d/2$ or $a \ll d$. In these cases, experiments which
measure $\langle E^2 \rangle$ are also detecting the mean energy
density, $\rho$.

\subsection{Comparison with the Model of Ford, Grove, and Ottewill}
\label{sec:FGO}

In this subsection, we will compare some aspects of our model with
the spin model of Ford, Grove, and Ottewill~\cite{FGO}, which was
summarized in Sect~\ref{sec:intro}. Ford {\it et al} assumed plane
wave modes, so $\rho = \langle E^2 \rangle$, whereas we use cavity
modes for which $\rho \not= \langle E^2 \rangle$, except in special
cases. Another difference is that the response of the spin system,
measured by a quantity analogous to our $\Delta P_2$,
seems to track the energy density whenever the system is far from
resonance, either $\omega \gg \Omega$ or  $\omega \ll \Omega$, where
$\omega$ is the frequency of the excited mode of the photon field 
and $\Omega$ is the resonant frequency of the spin system.  
By contrast, 
 we found a correlation of $\langle {\hat E}^2\rangle$ with
$P_2/P_2 (0)$ only near resonance, and a correlation with $\Delta P_2$
only far below resonance.
  This difference can be traced to the use
of adiabatic switching in Ref.~\cite{FGO}. 
At $t=t_0$, the spin system is assumed to be 
in the lower energy state, and then one takes the limit  $t_0
\rightarrow -\infty$, and averages oscillating quantities. In this
limit, for example
 ${\rm sin}^2 [(1/2)(\Omega - \omega) \,(t-t_0)] \rightarrow 1/2$, if
 $\Omega \neq \omega$. 
This is used to go from Eq. (4.8) to Eq. (4.9) in Ref.~\cite{FGO}. 
After this is done, it is not possible to let 
$\omega \rightarrow \Omega$. For example, the quantity 
\begin{equation}
\frac{{\rm sin}^2 [(1/2)(\Omega - \omega) \,(t-t_0)]}{{(\Omega-\omega)}^2} \,\, \rightarrow
 \left \{ 
\begin{array}{l}
\frac{1}{4} {(t-t_0)}^2 \,, {\rm for} \,\, \omega \rightarrow \Omega, \,\, {\rm with} \,\, t_0 
\,\, {\rm fixed, \, as\, in\, our\, case} \\
\frac{1}{2} {(\Omega-\omega)}^{-2}\,, {\rm adiabatic \,\,switching} \,.
\end{array}
\right.
\end{equation}
This describes a system which was prepared in the lower state and then coupled to the 
radiation field in the distant past. In this case, one obtains
expressions such as Eq.~(4.12) in Ref.~\cite{FGO},
with factors of $1/{(\Omega-\omega)}^{2}\,,1/ {(\Omega+\omega)}^{2}$, and $1/(\Omega^2-\omega^2)$, 
which are proportional to $\langle E^2 \rangle$ (or to the energy density, $\rho$), only if 
$\omega \gg \Omega$ or $\omega \ll \Omega$.

The adiabatic switching assumption may be appropriate for a system
coupled to photons in a plane wave mode, but does not apply to our
model. The entrance of the atoms into a cavity is more accurately
described by the sudden approximation employed here.

\section{Some Numerical Estimates}
\label{sec:NE}

\subsection{Criteria to be Fulfilled}

In this section, we will analyze the feasibility of observing the suppression of the 
de-excitation probability. We consider an atom passing through a cavity with a single mode 
excited.
Let us first list the desired criteria for a measurement of $P_2$. We wish 
to impose the following conditions:

\begin{enumerate}
\item[(1)] $\omega \approx \Delta \varepsilon$. This will allow us to
  apply Eq.~(\ref{eq:P2/P20_1}), and have a situation where the
  de-excitation probability tracks the mean squared electric field.

\item[(2)] It will be convenient to consider Rydberg atoms, which are
  often used in cavity QED experiments. Their transition frequencies
  are typically in the microwave range. Consider the specific case of
the transition between the $n=51$ and the $n=50$ energy 
levels,  with a transition frequency of $f = 51.1\,{\rm GHz}$  and wavelength 
$\lambda \approx 6\, {\rm mm}$~\cite{HR}.

\item[(3)] The excited cavity mode is the lowest mode, $\omega = \pi
  \sqrt{1/a^2+1/d^2}$, so
\begin{equation}
f_{cavity}=\frac{\omega}{2 \pi} = \frac{c}{2} \sqrt{\frac{1}{a^2}+\frac{1}{d^2}} \,.
\end{equation}
This requires the two longer dimensions of the cavity, $a$ and $d$,
each to be at least $3\, {\rm mm}$.

\item[(4)] For the atom to fit in the cavity, we need the smallest dimension of the cavity, $b$, to be 
larger than the size of the atom. The size of a Rydberg atom with
$n=50$ is about 100 nm. So we need
\begin{equation}
b > 100\, {\rm nm} = 0.1 \,\mu {\rm m} \,.
\end{equation}

\item[(5)] We want the travel time for the atom to traverse the cavity dimension $b$, 
to be no more than the time interval in which   $\langle {\hat E}^2
\rangle < 0$, which is a fraction of the period of the cavity mode.

 \item[(6)] To avoid the complications of relativity, and so that the atom sees a small Doppler shift 
 of the cavity frequency, we want
 \begin{equation}
\frac{v}{c} \ll 1 \,.
\end{equation}

\item[(7)]We would ideally like the lifetime, $\tau$, of the excited state to be roughly equal 
to the transit time of the atom across the cavity.
For the Rydberg atoms under consideration ($n=50$), $\tau \approx 3.6 \times 10^{-2}$ sec.
This would make $P_2(0)$ of order unity and maximize the effect we
are studying.
\end{enumerate}

\subsection{Order-of-Magnitude Estimates}
\label{sec:OME}

Let us consider the case of the vacuum plus two photon state discussed
in Sect.~\ref{sec:0p2}. Such a state might be created by a 
process which has a finite probability to emit a pair of photons into
the lowest mode of the cavity. We have seen that in this state
it is possible to have  ${P_2}/{P_2 (0)} \approx 0.55$, so there is a
significant reduction in the decay rate. The duration of the interval
of negative mean squared electric field is, from the discussion
at the end of Sect.~\ref{sec:0p2},
\begin{equation}
\Delta t \approx \frac{1}{6\,f} \approx 3 \times 10^{-11} {\rm s}\,.
\end{equation}
Let us choose the smallest dimension of the cavity to be 
\begin{equation}
b \approx1 \,\mu {\rm m} \,,
\end{equation}
which easily satisfies criterion (4). This requires that the speed 
of the atom be 
\begin{equation}
v > \frac{b}{\Delta t} \approx  3 \times 10^{5} {\rm m/s} \,.
\end{equation}
This can be satisfied by non-relativistic speeds compatible with
criterion (6). 

The one criterion listed above which cannot be satisfied is (7),
as the atomic lifetime is necessarily long compared to the transit
time. This means that each atom has a probability of less than
$10^{-9}$ of decaying while transiting the cavity. However, this need
not be a serious problem if a sufficiently large flux of atoms can
be used.

\section{Summary and Discussion}
\label{summary}
In this paper, we have considered the effects of vacuum fluctuation suppression on the de-excitation 
probabilities of atoms. This suppression occurs in quantum states of the radiation field in which 
the renormalized expectation value of the square of the electric field
operator is periodically negative. Examples of such states include the
vacuum plus two photon state and  the single-mode squeezed vacuum
state. Both of these are examples of quantum states exhibiting
sub-vacuum effects, such as negative energy density or negative mean
squared electric field. 

We have treated a model detector which consists of an atom in an
excited state traversing a cavity containing photons in a
non-classical state.
We calculated the ratio, $P_2/P_2(0)$, of the de-excitation
probability in 
an arbitrary single-mode cavity field 
quantum state to the same probability in the vacuum state. 
This calculation assumed that the atom and field are near 
resonance, with the transition frequency approximately equal to the lowest cavity frequency, so that 
the atom had only one available decay mode.  Our results showed that near resonance, the ratio 
$P_2/P_2(0)$ tracks  $\langle E^2 \rangle$. Hence $P_2/P_2(0)$ will have its 
minimum value when $\langle E^2 \rangle$ is maximally negative. In
certain cases, this is also when the energy density is maximally negative.

We also calculated the difference, $\Delta P_2 = P_2 - P_2(0)$, of the de-excitation probability 
between our arbitrary state and the vacuum state, where  the effects of the non-excited 
modes would cancel out. In this case $\Delta P_2 = P_2 - P_2(0)$ is proportional to $\langle E^2 \rangle$ 
far below resonance, but not in other cases.
 In particular, only when the lowest cavity mode frequency was much less than 
the transition frequency of the atom did we find that $\Delta P_2<0$
when $\langle E^2 \rangle<0$. In this case, we were able to derive a
quantum inequality lower bound on $\Delta P_2$.
 
Finally, we discussed the plausibility of an experiment involving
Rydberg atoms to detect $\langle E^2 \rangle<0$ and possibly negative
energy density. Although challenging, such an experiment may be
possible.

\begin{acknowledgments}
We would like to thank Ben Varcoe for valuable discussions.
This work was supported in part by the National
Science Foundation under Grants PHY-0555754 and PHY-0652904.
\end{acknowledgments}

\appendix

\section{Bound on $\Delta P_2$ }

Here we will derive a quantum inequality bound on the amount by which
the de-excitation probability may be suppressed by quantum field effects.
From Eq.~(\ref{eq:expE2}), we have
\begin{equation}
\langle \gamma_i |  {\hat E}^2 ({\bf x_0}, t) | \gamma_i \rangle  =
  {f^2} ({\bf x_0})\,\Big[2 \,   \langle n \rangle \,
+ 2\, \sum_{n=0}  \sqrt{(n+2) (n+1)} \,  \,  Re(c_n \, {c^*}_{n+2} \, e^{2i \omega t}) \Big] \,.
\label{eq:expE2_2}
\end{equation}
Let us examine the term in brackets. We can always choose the phases
of the $c$'s so that 
the magnitude of the second term will be largest at $t=0$. Let us call
the smallest possible value of the bracketed 
term $S$, i.e., 
\begin{equation}
S =2\, \langle n \rangle \,
+ 2\, \sum_{n=0}  \sqrt{(n+2) (n+1)} \,  \,  Re(c_n \, {c^*}_{n+2} )  \,.
\label{eq:defS}
\end{equation}
Following the argument given on p. 230 of Ref.\cite{F78}, we will prove that $S$, and thus 
$\langle E^2 \rangle$, are bounded from below. Since $S$ also appears in our expressions, 
Eq.~(\ref{eq:P2/P20_1}) and Eq.~(\ref{eq:DeltaP2}), we can show that these quantities are 
bounded below as well.
Using the fact that 
\begin{equation}
 \langle n \rangle \,
=  \sum_{n=0}^{\infty} \,n\, {|c_n|}^2  \,,
\label{eq:sumn}
\end{equation}
we may expand $S$ as follows,
\begin{equation}
S =  2 \,\sum_{n=0}^{\infty} \,n \,  {|c_n|}^2\,
+  \sum_{n=0}  \sqrt{(n+2) (n+1)} \,  \,  c_n \, {c^*}_{n+2}  + 
\sum_{n=0}  \sqrt{(n+2) (n+1)} \,  \,  {c^*}_n \, c_{n+2}  \,.
\label{eq:S_2}
\end{equation}
Rewriting the right-hand side of Eq.~(\ref{eq:sumn}), we have 
\begin{eqnarray}
\sum_{n=0}^{\infty} \,n\, {|c_n|}^2 & =&  \, \sum_{n=2}^{\infty}  (n-2) \,{|c_{n-2}|}^2
\nonumber \\
 & =&  \,  \sum_{n=1}^{\infty}  (n-1) \,{|c_{n-2}|}^2 - \sum_{n=2}^{\infty}  {|c_{n-2}|}^2
\nonumber \\
& =&  \,-1+ \sum_{n=1}^{\infty}  (n-1) \,{|c_{n-2}|}^2  \,.
\end{eqnarray}
Note also that 
\begin{eqnarray}
\sum_{n=0}^{\infty} \sqrt{(n+1)(n+2)} \,\, c_n \, {c^*}_{n+2}  
& = &  \, \sum_{\ell=2}^{\infty}  \sqrt{\ell -1} \, \sqrt{\ell}  \,\, c_{\ell-2} \, {c^*}_{\ell} 
\nonumber \\
 & =&  \,  \sum_{n=1}^{\infty} \sqrt{n(n-1)} \,\, c_{n-2} \, {c^*}_{n} \,.
\end{eqnarray}
To get the first equality, we let $n+2 \rightarrow \ell$; to get the second equality, we relabeled 
$\ell \rightarrow n$, and used the fact that the $n=1$ term does not contribute anything to the sum.
If we now substitute these expressions into Eq.~(\ref{eq:S_2}), and relabel appropriately, 
we can write $S$ as
\begin{eqnarray}
S &=& -1 + \sum_{n=1}^{\infty} \Big[ \,n \,  {|c_n|}^2  +   (n-1) \,{|c_{n-2}|}^2 
 + \sqrt{n(n-1)} \,\, c_{n-2} \, {c^*}_{n} +\sqrt{n(n-1)} \,\, {c^*}_{n-2} \, c_{n} \Big]
\nonumber \\
&=& -1+\sum_{n=1}^{\infty}{ |\sqrt{n} \, c_n +\sqrt{n-1} \,c_{n-2}|}^2 \,.
\end{eqnarray}
Hence $S \geq -1$, and so we have that 
\begin{equation}
\langle E^2 \rangle \geq -{f^2} ({\bf x_0}) \,, 
\label{eq:E2bound}
\end{equation}
and when $\omega \ll \varepsilon$, from Eq.~(\ref{eq:DeltaP2}), 
\begin{equation}
\Delta P_2 
\geq - \frac{2 \, {|\langle  \psi_2 | d_y |  \psi_1 \rangle|}^2 \, {f^2} ({\bf x_0}) }
{{ \Delta \varepsilon}^2} \,.
\label{eq:DeltaP2_QI}
\end{equation}


\begin{thebibliography}{28}  
\bibitem{EGJ} H. Epstein, V. Glaser, and A. Jaffe, Nuovo Cim. {\bf 36}, 1016 (1965).  

\bibitem{C} H.B.G. Casimir, Proc. K. Ned. Akad. Wet. {\bf B51}, 793
  (1948); 
 L.S. Brown and G.J. Maclay, Phys. Rev. {\bf 184}, 1272 (1969).   

\bibitem{SY} R. E. Slusher, L. W. Hollberg, B. Yurke, J. C. Mertz, and 
 J. F. Valley, Phys. Rev. Lett. {\bf 55}, 2409, (1985).  

 \bibitem{F78} L. H. Ford, Proc. Roy. Soc. Lond. {\bf A364}, 227 (1978).     
 
 \bibitem{F91} L. H. Ford, Phys. Rev. {\bf D43}, 3972 (1991).     
 
  \bibitem{LF100} L.H. Ford, "Spacetime in Semiclassical Gravity", in 
 {\it 100 Years of Relativity - Space-time Structure: Einstein and
   Beyond}, 
 edited by A. Ashtekar, (World Scientific, Singapore, 2006), gr-qc/0504096. 
  
\bibitem{TRMGM} T.A. Roman, ``Some Thoughts on Energy Conditions and
  Wormholes'',
 in {\it Proceedings of the Tenth Marcel Grossmann Meeting on General
   Relativity}, 
 edited by S.P. Bergliaffa and M. Novello, (World Scientific, Singapore,  2006), gr-qc/0409090.  

 \bibitem{CJF} C.J. Fewster, ``Energy inequalities in quantum field
   theory'', 
 in {\it XIVth International Congress on Mathematical Physics},
 edited by 
J.C. Zambrini (World Scientific, Singapore, 2005),  see updated version, math-ph/0501073.  

 \bibitem{M} P. Marecki, Phys. Rev. {\bf A 66}, 053801 (2002), quant-ph/0203027. 
  
 \bibitem{FGO} L.H. Ford, P.G. Grove, and A.C. Ottewill, Phys. Rev. D {\bf 46}, 4566 (1992). 

\bibitem{DO}  P.C.W. Davies, A.C. Ottewill, 
Phys. Rev. {\bf D65}, 104014 (2002), gr-qc/0203003.

\bibitem{Grove} P.G. Grove, Class. Quantum Grav. {\bf 5}, 1381 (1988).

 \bibitem{MS} P. Marecki and N. Szpak,  Ann. Phys. (Leipzig) 14, 428 (2005), quant-ph/0407186.
 
 \bibitem{GK} C.C. Gerry and P.L. Knight, {\it Introductory  Quantum
     Optics},
 (Cambridge University Press, Cambridge, 2005), p. 79 and Sec. 4.5.  


\bibitem{GO} N Graham and K. Olum, Phys. Rev. D {\bf 72},  025013 (2005), hep-th/0506136.

\bibitem{FOP} C. Fewster, K. Olum, and M. Pfenning,  Phys. Rev. D {\bf 75}, 025007 (2007), gr-qc/0609007.

 \bibitem{BFR} A. Borde, L.H. Ford, and T.A. Roman,  Phys. Rev. {\bf D65}, 084002 (2002), gr-qc/0109061. 

\bibitem{BB} M. Babiker and G. Barton,  Proc. Roy. Soc. Lond. {\bf
  A326}, 255 (1972); 277 (1972). 

\bibitem{HR}  S. Haroche and J-M. Raimond, {\it Exploring the Quantum: Atoms, Cavities, and Photons}, 
(Oxford University Press, New York, 2006), Chap. 5.

\bibitem{Caves} C. M. Caves, Phys. Rev. D \textbf{23}, 1693 (1981).
 
 \bibitem{Jackson} J.D. Jackson, {\it Classical Electrodynamics}, (3rd Edition, John Wiley \& Sons, Inc., 
 New York, 1999), Chap. 8.
 
 


   \end{thebibliography}
\end{document}